\documentclass[
superscriptaddress,
nofootinbib,
twocolumn,
 amsmath,amssymb,
 aps,
prl,
notitlepage
]{revtex4-1}

\usepackage[pdftex=true,linktoc=page]{hyperref} 
\usepackage{dcolumn}
\usepackage{bm}
\usepackage{epsfig}
\usepackage{graphicx}
\usepackage{amsfonts}
\usepackage[figuresright]{rotating}
\usepackage{psfrag}
\usepackage{float} 
\usepackage{footmisc}
\usepackage{caption}
\usepackage{fltpage}

\newcommand{\be}{\begin{equation}}
\newcommand{\ee}{\end{equation}}
\newcommand{\ba}{\begin{array}}
\newcommand{\ea}{\end{array}}
\newcommand{\bqa}{\begin{eqnarray}}
\newcommand{\eqa}{\end{eqnarray}}

\begin{document}

\title{Absolute magnetometry based on quantum beats in diamond nitrogen-vacancy centers}
\author{Kejie Fang}
\affiliation{Department of Physics, Stanford University, Stanford,
California 94305, USA}
\author{Victor M. Acosta}
\email{victor.acosta@hp.com}
\affiliation{Hewlett-Packard Laboratories, 1501 Page Mill Rd., Palo Alto, California 94304, USA}
\author{Charles Santori}
\affiliation{Hewlett-Packard Laboratories, 1501 Page Mill Rd., Palo Alto, California 94304, USA}
\author{Zhihong Huang}
\affiliation{Hewlett-Packard Laboratories, 1501 Page Mill Rd., Palo Alto, California 94304, USA}
\author{Kohei M. Itoh}
\affiliation{Graduate School of Fundamental Science and Technology, Keio University, Yokohama 223-8522, Japan}
\author{Hideyuki Watanabe}
\affiliation{Diamond Research Laboratory, National Institute of Advanced Industrial Science and Technology (AIST), Tsukuba Central 2-13,
1-1-1, Umezono, Tsukuba, Ibaraki 305-8568, Japan}
\author{Shinichi Shikata}
\affiliation{Diamond Research Laboratory, National Institute of Advanced Industrial Science and Technology (AIST), Tsukuba Central 2-13,
1-1-1, Umezono, Tsukuba, Ibaraki 305-8568, Japan}
\author{Raymond G. Beausoleil}
\affiliation{Hewlett-Packard Laboratories, 1501 Page Mill Rd., Palo Alto, California 94304, USA}

\begin{abstract}
We demonstrate an absolute magnetometer immune to temperature fluctuation and strain inhomogeneity, based on quantum beats in the ground state of nitrogen-vacancy centers in diamond. We apply this technique to measure low-frequency magnetic field noise using a single nitrogen-vacancy center located within 500 nm of the surface of an isotopically-pure (99.99\% $^{12}$C) diamond. The photon-shot-noise limited sensitivity achieves 38 nT/$\sqrt{\textrm{Hz}}$ for 4.45 s acquisition time, a factor of $\sqrt{2}$ better than the implementation which uses only two spin levels. For long acquisition times ($>$10 s), we realize up to a factor of 15 improvement in magnetic sensitivity, which demonstrates the robustness of our technique against thermal drifts. Applying our technique to nitrogen-vacancy center ensembles, we eliminate dephasing from longitudinal strain inhomogeneity, resulting in a factor of 2.3 improvement in sensitivity.  

\end{abstract}

\maketitle
Negatively-charged nitrogen-vacancy (NV) centers in diamond have become an attractive candidate for solid-state magnetometry with high sensitivity and nanoscale resolution \cite{lukin1, lukin2,degen,wrachtrup1}, due to their long coherence time \cite{wrachtrup2} and near-atomic size. The principle of NV-based magnetometry is detection of the Zeeman shift of the ground-state spin levels. Usually, two spin levels are utilized and the presence of a magnetic field induces an a phase shift in the spin coherence which can be detected optically \cite{odmr}. This scheme works well for AC (kHz-MHz) magnetometry \cite{wrachtrup2} and relatively low sensitivity DC field measurements \cite{lukin2, wrachtrup1,rondin, maletinsky, dcvictor,Pham}. Sensors based on this technique are being developed for applications ranging from neuroscience \cite{Pham, Hall}, cellular biology \cite{wrachtrup1, McGuinness}, superconductivity \cite{Bouchard}, and nano-scale magnetic resonance imaging \cite{Grinolds}.

Recently, it was discovered that the zero field splitting of the NV center ground state is temperature \cite{dDdT} and strain dependent \cite{efield}. Consequently, a magnetometer using two spin levels is subject to temperature fluctuation and strain inhomogeneity (if using an NV ensemble). This limits the magnetometer sensitivity \cite{dDdT, toyli} (for example temperature fluctuations of 0.01 $^\circ C$ lead to fluctuations in the magnetometer reading of $\sim$30 nT) and also has implications for quantum information processing \cite{maurer}.

In this Letter, we overcome these issues by exploiting the full spin-1 nature of the NV center \cite{lukin1,togen1,pulsecpt,reinhard} to observe quantum beats \cite{qb1,qb2} in the ground state with a beat frequency given only by the external magnetic field and fundamental constants. We use a single tone microwave field, which transfers all the population into a ``bright'' superposition of the $m_s=\pm 1$ levels. This technique enables measurement of weak magnetic fields at the nanometer scale over a broad range of frequencies. 

Quantum beating is a phenomenon of the time evolution of a coherent superposition of non-degenerate energy eigenstates at a frequency determined by their energy splitting. It has wide applications in atomic spectroscopy \cite{qb3,qb4}, and vapor-cell magnetometry \cite{qbmag}. The phenomenon is closely related to coherent population trapping, which has been demonstrated in many different systems including quantum dots \cite{cptqd1,cptqd2}, superconducting phase qubits \cite{Kelly} and NV centers \cite{santoricpt, togen1}.

\begin{figure}[H]
\centering\epsfig{file=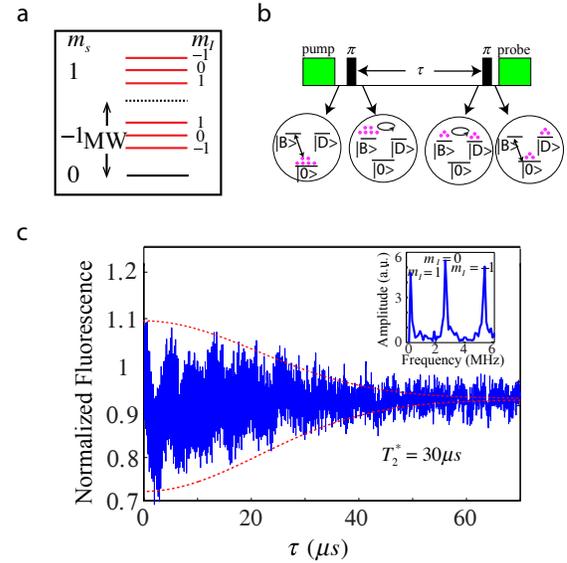,clip=1,width=1\linewidth,angle=0}\vspace{-2cm}\hspace{0.3cm}
\caption{\textbf{a} A single-tone microwave pulse interacts with all NV center ground-state sublevels. \textbf{b} Ramsey type magnetometry using quantum-beats between $|B\rangle$ and $|D\rangle$. \textbf{c} Ramsey fringes using the protocol in \textbf{b}. The fitted decay envelope yields $T_2^\star=30(1)$ $\mu$s. Inset: Absolute value of the fourier transform of the Ramsey fringes. The three peaks correspond to the three hyperfine resonances between $m_s=\pm 1$ levels.}
\label{scheme}
\end{figure}

Our quantum-beats magnetometer utilizes a linearly-polarized microwave field with frequency $f$ and transverse amplitude $B_{\textrm{MW}}$ (perpendicular to NV axis)  interacting with the $S=1$ NV ground state (Fig. \ref{scheme}a). The Hamiltonian describing this interaction is
\bqa \label{ac} H/h&=&\nu_+|1\rangle\langle1|+\nu_-|\mbox{-}1\rangle\langle\mbox{-}1|\\\nonumber&&-\Omega_{R0}\textrm{cos}(2\pi ft)(|1\rangle\langle0|+|0\rangle\langle1|+|\mbox{-}1\rangle\langle0|+|0\rangle\langle\mbox{-}1|) ,\eqa where $\Omega_{R0}=g_e\mu_B B_{\textrm{MW}}/\sqrt{2}$ is the undressed Rabi frequency, $\mu_B=13.996$ GHz/T is the Bohr magneton, $g_e=2.003$ is the NV electron $g-$factor, $h$ is Planck$^\prime$s constant, and $\nu_{\pm}$ is the transition frequency between $|0\rangle$ and $|\pm 1\rangle$. Here $|m_s\rangle$ denotes the ground state with spin projection $S_z=m_s$. From Eq. (\ref{ac}), we see that the microwave field only drives transitions between $|0\rangle$ and a certain superposition of $|\pm 1\rangle$, called the bright state, $|B\rangle=(|1\rangle+|\mbox{-}1\rangle)/\sqrt{2}$. The orthogonal superposition, $|D\rangle=(|1\rangle-|\mbox{-}1\rangle)/\sqrt{2}$, does not interact with the microwave field and is therefore called the dark state. If $\Omega_{R0}\gg |f-\nu_\pm|$, then Eq. (\ref{ac}) describes the Rabi oscillation between $|0\rangle$ and $|B\rangle$, and the precession between $|B\rangle$ and $|D\rangle$ due to the difference of $\nu_\pm$ can be ignored (see Supplementary information).

Our proposed magnetometer works in the weak field and weak transverse strain regime when the transition frequencies are \cite{efield} \be\label{eigenenergy}  \nu_{\pm}\approx D+d^\parallel\epsilon_z\pm( g_e\mu_BB_z+A_{||}m_I),\ee  where $D\approx$ 2.87 GHz is the ground state zero field splitting, $d^\parallel$ is the axial ground-state electric dipole moment, $\epsilon_z$ is the axial electric field (crystal strain), $A_{||}=-2.16$ MHz \cite{Smeltzer} is the parallel hyperfine coefficient, and $m_I$ is the spin projection of the $^{14}$N nucleus ($I=1$). This corresponds to the limit $|\vec B|\ll D/g_e\mu_B\approx 0.1$ T, $|g_e\mu_BB_z+A_\parallel m_I|\gg  (g_e\mu_BB_\perp+A_\perp m_I)^2/D$, and $|g_e\mu_BB_z+A_\parallel m_I|\gg|d_\perp \epsilon_\perp|$, where $d_\perp$ and $\epsilon_\perp$ are the non-axial ground-state electric dipole moment and electric field respectively, and $A_\perp=-2.7$ MHz \cite{felton} is the perpendicular hyperfine coefficient. Note the last condition does not set a minimum detectable magnetic field, since for usual diamond samples $d_\perp \epsilon_\perp$ is in the kHz range, and therefore the condition is always satisfied for at least two nuclear sub-levels.

The experimental scheme, based on Ramsey interferometry, is schematically shown in Fig. \ref{scheme}b. A green laser pulse initializes the NV electronic spin into $|0\rangle$. A single-tone $\pi$ pulse of sufficient spectral width is then applied to transfer the spin into $|B\rangle$. After a free evolution time $\tau$, the state becomes \bqa \label{bd} |\psi(t)\rangle&=&(e^{2\pi i\nu_+\tau}|1\rangle+e^{2\pi i \nu_-\tau}|\mbox{-}1\rangle)/\sqrt{2}\nonumber\\&=&e^{\pi i( \nu_++ \nu_-)\tau}\{\textrm{cos}[\pi( \nu_+-\nu_-)\tau]|B\rangle\nonumber\\&&+i\textrm{sin}[\pi( \nu_+- \nu_-)\tau]|D\rangle\}.\eqa  We see from Eq. (\ref{bd}) a population evolution between $|B\rangle$ and $|D\rangle$ with a beating frequency $\nu_+-\nu_-$. Then, a second $\pi$ pulse which is phase-coherent with the first $\pi$ pulse projects the population in $|B\rangle$ back to $|0\rangle$, while the population in $|D\rangle$ is trapped. A final green laser pulse induces the normalized, ensemble-averaged fluorescence signal $P(\tau)\propto(1+F(\tau)\textrm{cos}(4\pi(g_e\mu_B B_z+A_\parallel m_I)\tau))/2$, where, for Gaussian decay, $ F(\tau)\propto e^{-(\tau/T_2^\star)^2}$ with $T_2^\star$ the dephasing time. By monitoring $P(\tau)$ for fixed $\tau\approx(2n+1)/(8(g_e\mu_B B_z+A_\parallel m_I))$ where $n$ is an integer (maximizing the slope of $P(\tau)$), we can measure changes in $B_z$. Since $P(\tau)$ only depends on fundamental constants and $B_z$, the quantum-beats magnetometer is immune to temperature fluctuation and strain inhomogeneity, and it is absolute without the need of calibration (see Supplementary information).

In comparison, previous magnetometry demonstrations \cite{lukin2,degen,wrachtrup1,wrachtrup2} used a large bias magnetic field such that coherence between $|0\rangle$ and only one of $|\pm1\rangle$ was selectively addressed. Broadband magnetometry was realized using Ramsey interferometry, which begins with a green laser pulse used to initialize the spin into $|0\rangle$, followed by a microwave $\pi/2$ pulse which creates the state $(|0\rangle+|1\rangle)/\sqrt{2}$. After a free evolution of time $\tau$, a second $\pi/2$ pulse is applied to project the state to $|0\rangle$, which is then read out optically. The resulting fluorescence signal is $P(\tau)\propto(1+F(\tau)\textrm{cos}(2\pi\delta \tau))/2$, where $\delta=| \nu_+-f|$. Since $\delta$ depends on both $D$ and $\epsilon_z$ (Eq. (\ref{eigenenergy})), the measurement suffers from the temperature dependence of $D$ \cite{dDdT} and inhomogenity in $\epsilon_z$ if using an NV ensemble.

Our experiments demonstrate that overcoming these constraints is critical for high-sensitivity measurement of low-frequency magnetic fields. We used an isotopically-purified $^{12}$C sample ([$^{12}$C]=99.99\%) \cite{toyo}  to study the temperature sensitivity of our quantum-beats magnetometer. The sample has a 500-nm thick isotopically-pure layer with [NV]$\approx 10^{11}$ cm$^{-3}$ grown on top of a naturally-abundant substrate with negligible NV density. Isotopically purified diamond samples are particularly appealing for quantum information and sensing applications due to the long spin dephasing times afforded by the nearly spinless carbon lattice \cite{wrachtrup2,maurer, toyo, purifiedsample,zhao,jelezko}. A homebuilt confocal microscope was used in the experiment. Light from a 532 nm laser ($\sim1.2$ mW) illuminated the sample through an oil-immersion objective with 1.3 numerical aperture, and the fluorescence was collected, spectrally filtered, and detected with an avalanche photodiode. Pump and probe durations were 2 and 0.3 $\mu$s, respectively. A 25 $\mu$m diameter copper wire was attached to the surface of the sample to provide square microwave field pulses with a Rabi frequency $\sim 20$ MHz.

We performed Ramsey interferometry on NV centers using both the typical 2-level scheme ($\{0,1\}$ basis) and quantum-beats detection scheme ($\{1,-1\}$ basis). A small bias field ($<$2 G) was applied. The spin coherence time $T_2^\star$ varies among NV centers in this sample and we chose one with relatively long $T_2^\star$. The measured $T_2^\star$ for $\{0,1\}$ basis and $\{1,-1\}$ basis is $62(2)$ $\mu$s and $30(1)$ $\mu$s, respectively (Fig. \ref{scheme}c). 

\begin{figure}[H]
\centering\epsfig{file=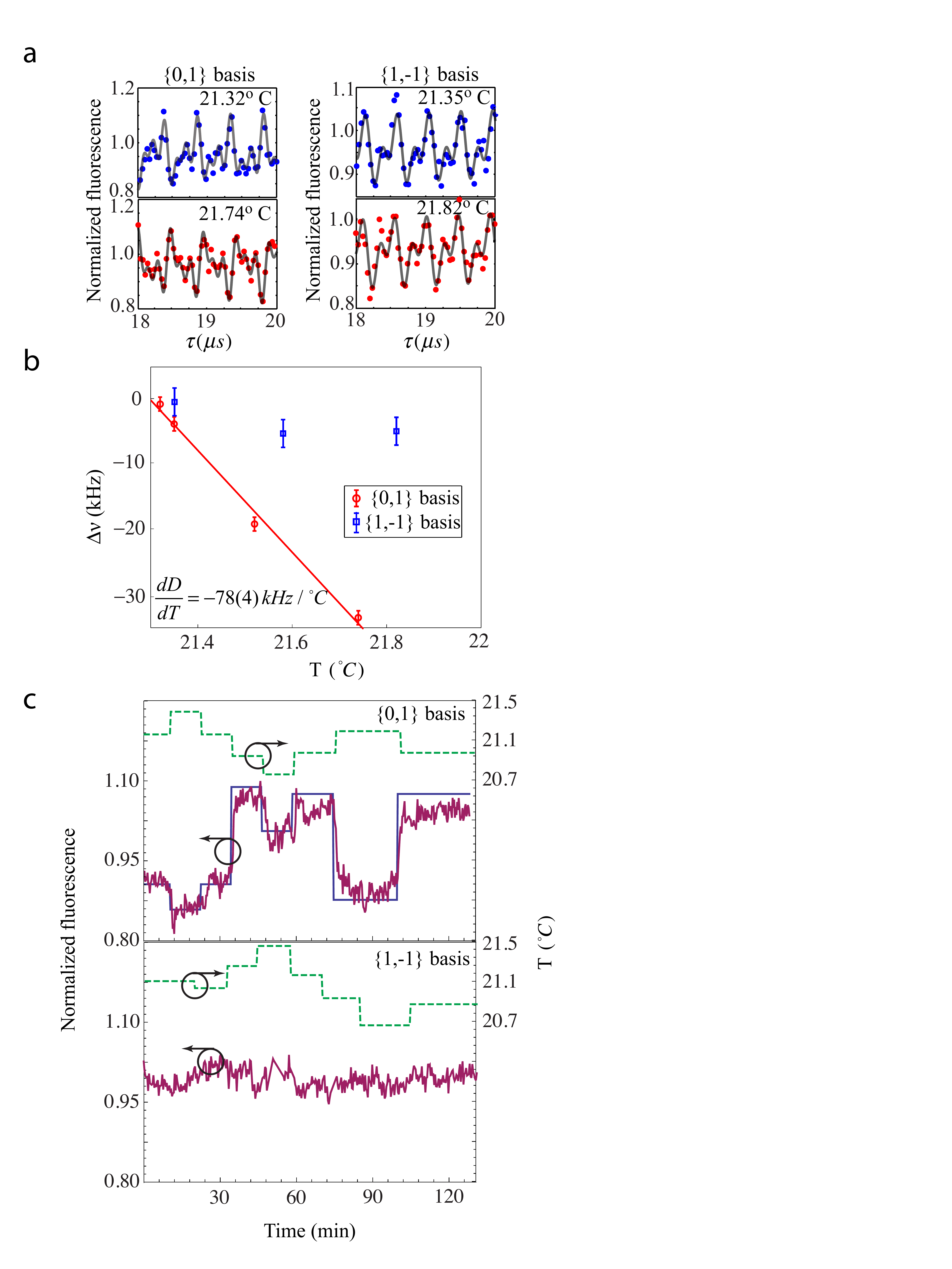,clip=1,width=1.5\linewidth,angle=0}\vspace{0.3cm}\hspace{0.3cm}
\caption{\textbf{a} Ramsey fringe of the $\{0,1\}$ basis (left panel) and $\{1,-1\}$ basis (right panel) for different temperatures. \textbf{b}  Zero field splitting $D$ dependence on temperature as measured in the $\{0,1\}$ basis (red). The $\{1,-1\}$ basis is immune to changing $D$ (blue). \textbf{c} Fluorescence level dependence on temperature for fixed delay time, 18.475 $\mu$s for the $\{0,1\}$ basis (upper panel) and 19.720 $\mu$s for the $\{1,-1\}$ basis (lower panel). Red curves are measured data. Dashed lines indicates the temperature change. Solid lines are the calculated fluorescence level for the $\{0,1\}$ basis.  }
\label{Tdependence}
\end{figure}

Using a thermoelectric element, we varied the temperature of the diamond sample, and performed $P(\tau)$ measurements using both the $\{0,1\}$ and $\{1,-1\}$ bases. The results are plotted in Fig. \ref{Tdependence}a. We see a clear temperature dependence in the shape of Ramsey fringes for the $\{0,1\}$ basis which is not present in the $\{1,-1\}$ basis. We fit the data with a model containing three hyperfine levels $ P(\tau) =\sum_{i=1}^3A_i\textrm{cos}(2\pi\nu_i\tau+\phi_i)+b$, where $A_i$, $\nu_i$ and $\phi_i$ are the amplitude, frequency and phase of the three hyperfine oscillations respectively, and $b$ is a constant. We used a global fit in which $\nu_i=\nu_{i,T_0}+\Delta \nu(T)$, and $A_i$, $\nu_{i,T_0}$, $\phi_i$ and $b$ are fixed for all the temperatures to fit for $\Delta \nu(T)$ (Fig. \ref{Tdependence}b). For the $\{0,1\}$ basis, assuming $\Delta \nu(T)=\Delta D(T)$, we find $dD/dT=-78(4)$ kHz/$^\circ C$, which is consistent with the previous report \cite{dDdT}. Finally, we fixed the delay time of the Ramsey interferometer and measured the fluorescence level as the temperature was varied. As shown in Fig. \ref{Tdependence}c, the fluorescence level in the $\{0,1\}$ basis changed significantly and can be well fitted with the parameters obtained from fitting the temperature dependence of the Ramsey curves. In comparison, the change of fluorescence level in the $\{1,-1\}$ basis is about a factor of 7 smaller and does not appear to be correlated with the changes in temperature.

Another advantage of working in the $\{1,-1\}$ basis is the improvement of the magnetometry sensitivity by a factor of $\sqrt{2}$. Consider the minimum detectable field, $b_{min}$, of a Ramsey-type magnetometer limited by quantum-projection fluctuations. It is determined by $b_{min}=\Delta N/(\Delta m_sg_e\mu_B|\partial N/\partial \nu|)$, where $N=\textrm{cos}(2\pi\nu \tau)$ is the probability distribution difference in the two levels, $\nu$ is the spin-precession frequency in the rotating frame, $\Delta N$ is the projection noise, and $\Delta m_s$ is the magnetic quantum number difference of the two levels. In both schemes we only measure the probability distribution in two levels, so we can represent the two-level system as a spin-$\frac{1}{2}$ system. Then the single-shot projection noise is $\Delta N=\langle \Delta \sigma_z\rangle=\sqrt{\langle \psi|\sigma_z^2|\psi\rangle-\langle \psi|\sigma_z|\psi\rangle^2}$, where $\sigma_z$ is the Pauli matrix and $\psi$ is the final state. As $\langle \psi|\sigma_z|\psi\rangle$ is simply the signal $N$, we have $\Delta N=|\textrm{sin}(2\pi\nu \tau)|$. Inserting this expression into the definition of $b_{min}$, we find $ b_{min}(\tau)=\frac{1}{2\pi\Delta m_s g_e\mu_B C(\tau)\tau}$ where $C(\tau)=e^{-(\tau/T_2^\star)^2}$ is the contrast decay due to Gaussian noise. For multiple measurements, $b_{min}$ can be improved by a factor of $\sqrt{T/\tau}$, where $T$ is the total measurement time. The quantum-projection-noise limited sensitivity for our magnetometer is thus defined as \be\label{stvt} \eta_{min}(\tau)\equiv b_{min}\sqrt{T}=\frac{1}{2\pi\Delta m_s g_e\mu_B e^{-(\tau/T_2^\star)^2}\sqrt\tau}. \ee The best sensitivity is achieved at $\tau=T_2^\star/2$. Although $T_2^\star$ in the $\{1,-1\}$ basis is half of that in $\{0,1\}$ basis, $\Delta m_s$ is twice as big, so $\eta_{min}$ is improved by a factor of $\sqrt{2}$ in $\{1,-1\}$ basis (see Supplementary information).

In our experiment, due to finite photon collection efficiency and imperfect spin-state readout, we can only measure the photon-shot-noise limited sensitivity (see Supplementary information). However, the $\sqrt{2}$ improvement of sensitivity in the $\{1,-1\}$ basis still persists since the contrast of $P(\tau)$ curves and photon collection efficiency are the same for the two bases. For $T=5$ s, the optimal measured sensitivity for the $\{1,-1\}$ basis and $\{0,1\}$ basis is 38(3) and 53(4) $\textrm{nT}/\sqrt{\textrm{Hz}}$, respectively. The ratio of the two sensitivities is $1.39(15)$, which is consistent with the theoretical value of $\sqrt{2}$. In comparison, the quantum-projection-noise limited sensitivity calculated using Eq. (\ref{stvt}) is 0.79 and 1.12 $\textrm{nT}/\sqrt{\textrm{Hz}}$ respectively (we used the conversion $\sqrt{\textrm{s}}\leftrightarrow\sqrt{2} /\sqrt{\rm Hz}$.). 

\begin{figure}[H]
\centering\epsfig{file=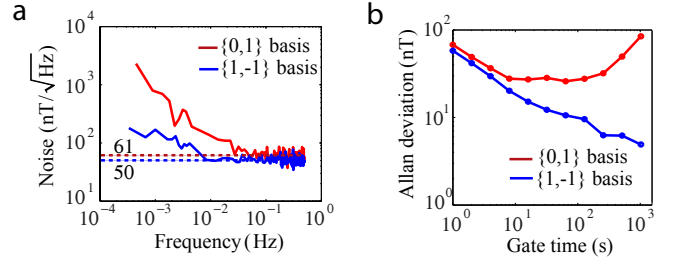,clip=1,width=1.2\linewidth,angle=0}\vspace{-8.3cm}\hspace{0.3cm}
\caption{\textbf{a}  Measured noise spectrum using the $\{0,1\}$ and $\{1,-1\}$ bases. Dashed lines are the noise floor near 1 Hz. \textbf{b} Allan deviation of the noise.   }
\label{sensitivity}
\end{figure}

We used both schemes to measure the real noise in the laboratory. Using $\tau=14.5$ $\mu$s (29 $\mu$s) for $\{1,-1\}$ ($\{0,1\}$) basis, we repeated the fluorescence measurement in 1-second intervals for 50 minutes, now without any active temperature control. As seen from Fig. \ref{sensitivity}a, for frequencies near 1 Hz, the noise floor measured by the $\{1,-1\}$ and $\{0,1\}$ bases is $50$ and $61$ $\textrm{nT}/\sqrt{\textrm{Hz}}$ respectively. If we ignore dead time from state preparation and readout, the sensitivity is 43 and 56 nT/$\sqrt{\rm Hz}$ respectively, consistent with the photon-shot-noise limit. For lower frequencies, the $\{0,1\}$ basis suffers more noise which is presumably due to laboratory temperature fluctuations. 

To further elucidate this effect, we analyzed the two-sample Allan deviation \cite{allan} of the fluorescence data, as shown in Fig. \ref{sensitivity}b. The Allan deviation in the $\{0,1\}$ basis at long gate time ($T_{\rm gate}$) levels off and even begins to increase at $T_\textrm{gate}\sim 100$ s, indicating that averaging the signal for a longer period of time no longer improves estimation of a static magnetic field. In contrast, the Allan Deviation continues to decrease for the $\{1,-1\}$ basis up to $T_\textrm{gate}\ge1000$ s, indicating this technique is suitable for distinguishing nT-scale static fields by using long integration times.

\begin{figure}[H]
\centering\epsfig{file=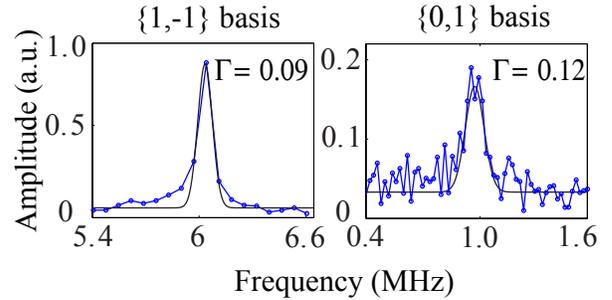,clip=1,width=1.2\linewidth,angle=0}\vspace{-9.3cm}\hspace{0.3cm}
\caption{Fourier transform of $P(\tau)$ in both bases for an ensemble of NV centers. Each peak corresponds to the $m_I=-1$ level. Black curves are Gaussian fits.  }
\label{strainpeak}
\end{figure}

Finally, we studied the effect of strain inhomogeneity on a magnetometer employing an ensemble of NV centers. We expect that, from Eq. (\ref{eigenenergy}), $P(\tau)$ measurement in the $\{1,-1\}$ basis is insensitive to strain inhomogeneity; however, there will be inhomogenous broadening, and consequently reduction of $T_2^\star$, in the $\{0,1\}$ basis due to variations in $\epsilon_z$ for each NV center in the ensemble. We used a sample with [$^{12}$C]=99.9\%, which was implanted with $10^{10}/\textrm{cm}^2$ $^{14}$N$^+$ at an energy of 20 keV and annealed at 875 $^\circ C$ for 2 hours, resulting in an NV density of $\sim 5/ \mu\textrm{m}^2$. The laser spot was defocused to illuminate a $\sim$2.5 $\mu$m diameter region and the optical power was increased to 30 mW to maintain constant intensity. A bias field $B_z\approx0.28$ G was applied along the [100] direction, such that NVs with different orientation experience the same $|B_z|$. For the $\{1,-1\}$ basis, we used microwave pulses with enough spectral width to cover all three hyperfine levels. For the $\{0,1\}$ basis, we detuned the microwave frequency and reduced the power to selectively address the $m_I=-1$ level. The peak corresponding to the $m_I=-1$ level of the Fourier transform of $P(\tau)$ is shown in Fig. \ref{strainpeak} for both cases. Gaussian fits revealed a full-width-at-half-maximum $\Gamma=0.09$ and 0.12 MHz for the $\{1,-1\}$ and $\{0,1\}$ bases, respectively. This indicates a factor of $\sim$2.7 increase in spin linewidth due to inhomogenous broadening in the $\{0,1\}$ basis, since for a single NV $\Gamma$ would be half of that of $\{1,-1\}$ basis. Accordingly, we estimate the longitudinal strain inhomegeneity in the detected region is $\sim100$ kHz. The result indicates a factor of 2.3 improvement of sensitivity for the quantum-beats magnetometer according to Eq. (\ref{stvt}). Our result also sheds light on other NV-ensemble applications such as quantum memories \cite{kubo, zhu} and frequency references \cite{hodges}.

In summary, we have demonstrated a magnetometer insensitive to temperature fluctuation and strain inhomogeneity based on quantum-beats in NV centers in diamond. The new method uses a similar pulse sequence and does not increase the complexity of the magnetometer.

{\bf Acknowledgement}
We thank D. Budker, P. R. Hemmer, K.-M. C. Fu, and T. Ishikawa for contributing valuable ideas during the conception of this experiment. K.F. acknowledges the support of S. Fan and Stanford Graduate Fellowship.

\newpage

\newpage

\section{Supplementary information}

\setcounter{figure}{0}
\setcounter{equation}{0}
\setcounter{page}{1}

\makeatletter
\renewcommand{\thefigure}{S\@arabic\c@figure}
\renewcommand{\theequation}{S\@arabic\c@equation}
\makeatother
\renewcommand{\bibnumfmt}[1]{[S#1]}
\renewcommand{\citenumfont}[1]{S#1}

\subsection{Detail analysis of possible limitations of our approach}

\subsubsection{Effect of imperfect $\pi$ pulse}

In this section, we analyse the influence of the imperfect $\pi$ pulse on the residual temperature-dependence in the $\{1,-1\}$ basis. If the $\pi$ pulse does not have enough spectral width, then there is some residual population in $|0\rangle$ after the $\pi$ pulse, which will evolve in the $\{0,1\}$ basis and thus suffers from the noise from temperature fluctuation. Below, we give an estimation of the residual population. The Hamiltonian (Eq. (1)) can be simplified under the rotating wave approximation and in the rotating frame to be time-independent, \bqa\label{rwa} H/h&=&(\nu_+-f)|\tilde1\rangle\langle\tilde1|+(\nu_--f)|\widetilde{\mbox{-}1}\rangle\langle\widetilde{\mbox{-}1}|\\\nonumber&&-\frac{\Omega_{R0}}{2}(|\tilde1\rangle\langle0|+|0\rangle\langle\tilde1|+|\widetilde{\mbox{-}1}\rangle\langle0|+|0\rangle\langle\widetilde{\mbox{-}1}|) ,\eqa where $|\tilde 1\rangle=e^{2\pi i ft}|1\rangle$ and $|\widetilde{\mbox{-}1}\rangle=e^{2\pi i ft}|\mbox{-}1\rangle$. If we define the bright state and dark state in the rotating frame as $|\tilde B\rangle=\frac{|\tilde1\rangle+|\widetilde{\mbox{-}1}\rangle}{\sqrt{2}}$ and $|\tilde D\rangle=\frac{|\tilde1\rangle-|\widetilde{\mbox{-}1}\rangle}{\sqrt{2}}$, Eq. (\ref{rwa}) can be rewritten in the basis $\{|0\rangle, |\tilde B\rangle, |\tilde D\rangle\}$ as \bqa  H/h&=&\frac{1}{2}(\nu_+-\nu_-)(|\tilde B\rangle\langle\tilde D|+|\tilde D\rangle\langle\tilde B|)\\\nonumber&&-\frac{\Omega_{R0}}{2}(|\tilde B\rangle\langle0|+|0\rangle\langle\tilde B|)\\\nonumber&& -\frac{1}{2}(\nu_++\nu_--2f)|0\rangle\langle 0|.\eqa

Writing an ansatz for the state \be |\Psi(t)\rangle=c_0(t)|0\rangle+c_B(t)|\tilde B\rangle+c_D(t)|\tilde D \rangle \ee and inserting it into the Schr\"{o}dinger$'$s equation, we get \bqa \label{s1}\frac{dc_0}{dt}=i\pi\Omega_{R0}c_B+i\pi(\nu_++\nu_--2f)c_0,\\\label{s2} \frac{dc_B}{dt}=i\pi\Omega_{R0}c_0-i\pi(\nu_+-\nu_-)c_D, \\ \label{s3} \frac{dc_D}{dt}=-i\pi(\nu_+-\nu_-)c_B.\eqa  Combining Eq. (\ref{s1}) and Eq. (\ref{s2}), we have \be\label{s4} \frac{d^2c_0}{dt^2}-i\pi(\nu_++\nu_--2f)\frac{dc_0}{dt}+\pi^2\Omega_{R0}^2c_0=\pi^2\Omega_{R0}(\nu_+-\nu_-)c_D.\ee When $\Omega_{R0}\gg |\nu_+-\nu_-|$, Eq. (\ref{s4}) can be solved using perturbation method by expanding the solution $c_0=c_0^{(0)}+c_0^{(1)}+\cdots$, where the superindex $k$ represents the order of $O((\frac{\nu_+-\nu_-}{\Omega_{R0}})^k)$. The zeroth order solution gives the Rabi oscillation between $|0\rangle$ and $|\tilde B\rangle$ with Rabi frequency $\Omega_R=\sqrt{\Omega_{R0}^2+(\nu_++\nu_--2f)^2/4}$. The first order correction gives the residual population in $|0\rangle$ after the imperfect $\pi$ pulse, which is $|c_0^{(1)}(\frac{1}{2\Omega_R})|^2\approx \frac{(\nu_+-\nu_-)^4}{4\Omega_R^4}$. This could be taken as a figure-of-merit to estimate the temperature dependence due to imperfect $\pi$ pulse in $\{1,-1\}$ basis. As an estimation, in our experiment, we have $\nu_+-\nu_-\approx 4$ MHz, $\Omega_R\approx 20$ MHz, $f\approx(\nu_++\nu_-)/2$. The residual population is thus $|c_0^{(1)}|^2\approx 4\times10^{-4}$.

\subsubsection{Magnetometer offset due to transverse strain}

The expression for the transition frequency $\nu_\pm$ (Eq. (2)) is modified when $|d_\perp \epsilon_\perp|\ll |g_e\mu_B B_z+A_\parallel m_I|$ is not satisfied (the other two conditions for Eq. (2) to hold is generally satisfied due to large $D$). In this case, \bqa\label{vmod}  \nu_{\pm}&=&D+d^\parallel\epsilon_z\pm \sqrt{(g_e\mu_B B_z+ A_{||}m_I)^2+(d^\perp\epsilon_\perp)^2}\\\nonumber &\approx& D+d^\parallel\epsilon_z\pm (g_e\mu_B B_z+A_{||}m_I+\frac{(d^\perp\epsilon_\perp)^2}{2(g_e\mu_B B_z+ A_{||}m_I)}),\eqa

Typical values for single NV centers in our sample are $d^\perp\epsilon_\perp\approx10$ kHz. Thus even for moderate values of $|B_z|\approx0.5~{\rm G}$, the shift is less than $100~{\rm Hz}$, corresponding to just a few nT.

\subsubsection{Temperature dependence due to transverse strain}
The temperature dependence of the transverse strain splitting was previously studied for ensembles in a variety of high-defect-density samples \cite{dDdTs}. In that work the term $d^\perp\epsilon_\perp$ was defined as $E$ and the fractional temperature dependence was determined to be $\frac{1}{E}\frac{dE}{dT}=1.4(3)\times10^{-4}~{\rm K^{-1}}$. Taking the case of $E\approx10~{\rm kHz}$ and $|B_z|\approx0.5~{\rm G}$, the temperature dependence of the extra term in Eq. (\ref{vmod}) $\frac{(d^\perp\epsilon_\perp)^2}{2(g_e\mu_B B_z+ A_{||}m_I)}$ is of order $10^{-2}~{\rm Hz/K}$. This is more than $6$ orders of magnitude smaller than $dD/dT$ and undetectable in our experiments.

\subsubsection{Temperature dependence of $g_e$ and $A_{\parallel}$}
To our knowledge there have been no published studies of the temperature dependence of the NV g-factor and hyperfine coupling coefficients. We note that the NV electron $g$ factor $g_e$ only differs by a tiny amount ($\sim0.03\%$) from that of a free electron spin due to the small spin-orbit interaction in diamond. As a result, we expect it is much less sensitive to lattice expansion.

Measurements of the NV $g$-factor, $g_e$ have been performed on ensembles at various temperatures from liquid helium \cite{takahashis} to room temperature \cite{feltons}. The reported values have been consistent, $g_e=2.0029$, to within the experimental uncertainty of $\sim 0.0002$ \cite{feltons}. As a rough approximation on an upper bound of the magnitude of the effect, we might assume that $g_e$ varies linearly by $.0004$ over the temperature interval from $4$ to $295$ K. In this case we find $\frac{dg_e}{dT}\approx10^{-6}$ K$^{-1}$. The overall temperature sensitivity of the magnetometer signal would be $\Delta m_s \mu_B |B_z|\frac{dg_e}{dT}\approx4~{\rm Hz/K}$ for $|B_z|\approx0.5~{\rm G}$ and $\Delta m_s=2$. This conservative bound is already more than $4$ orders of magnitude smaller than $dD/dT$ and could be further suppressed by operating at lower field.

Separate measurements of $A_{\parallel}$ for single NV centers have been performed at $8~{\rm K}$, $A_{\parallel}(8~{\rm K})=-2.169(7)$  \cite{robledos}, and room temperature $A_{\parallel}(295~{\rm K})=-2.162(2)$ \cite{Smeltzers}. The difference in the measured values is within the experimental error. A rough estimate can be obtained by taking a linear extrapolation, yielding $dA_{\parallel}/dT\approx24(25)~{\rm Hz/K}$. This is more than $3$ orders of magnitude smaller than $dD/dT$ and is not discernable in our experiments.

\subsection{$T_2^\star$ of $\{1,-1\}$ basis and $\{0,1\}$ basis}

The coherence in the $\{0,1\}$ basis can be characterized by the average of the phase factor as $L_{0,1}=\langle e^{i\phi(t)} \rangle$ subject to environment noise \cite{pulsecpts}, where $\phi(t)=\int (\nu_+-f)dt$. Then the coherence in the $\{1,-1\}$ basis can be characterized by the average of the phase factor as $L_{1,-1}=\langle e^{2i\phi(t)} \rangle$. The phase $\phi(t)$ is correlated to the environmental noise $S$ (for example, through Eq. (2) for magnetic noise), and thus the $\{1,-1\}$ basis suffers from twice as much noise as the $\{0,1\}$ basis. When the noise correlation time is much longer than the dephasing time $T_2^\star$, e.g. Gaussian noise, then $1/T_2^\star$ is proportional to the noise amplitude \cite{charlesbook}. As a result, in this case $T_2^\star(\{0,1\})=2T_2^\star(\{1,-1\})$, and we have a $\sqrt{2}$ improvement in sensitivity. On the other hand, if the noise correlation time is much shorter than the dephasing time $T_2^\star$, then $1/T_2^\star$ is proportional to the square of the noise amplitude \cite{charlesbook}, and thus $T_2^\star(\{0,1\})=4T_2^\star(\{1,-1\})$. Consequently, there is no sensitivity improvement using the $\{1,-1\}$ basis in a rapidly fluctuating environment.

\subsection{Photon-shot-noise limited sensitivity}

\begin{figure}[H]
\centering\epsfig{file=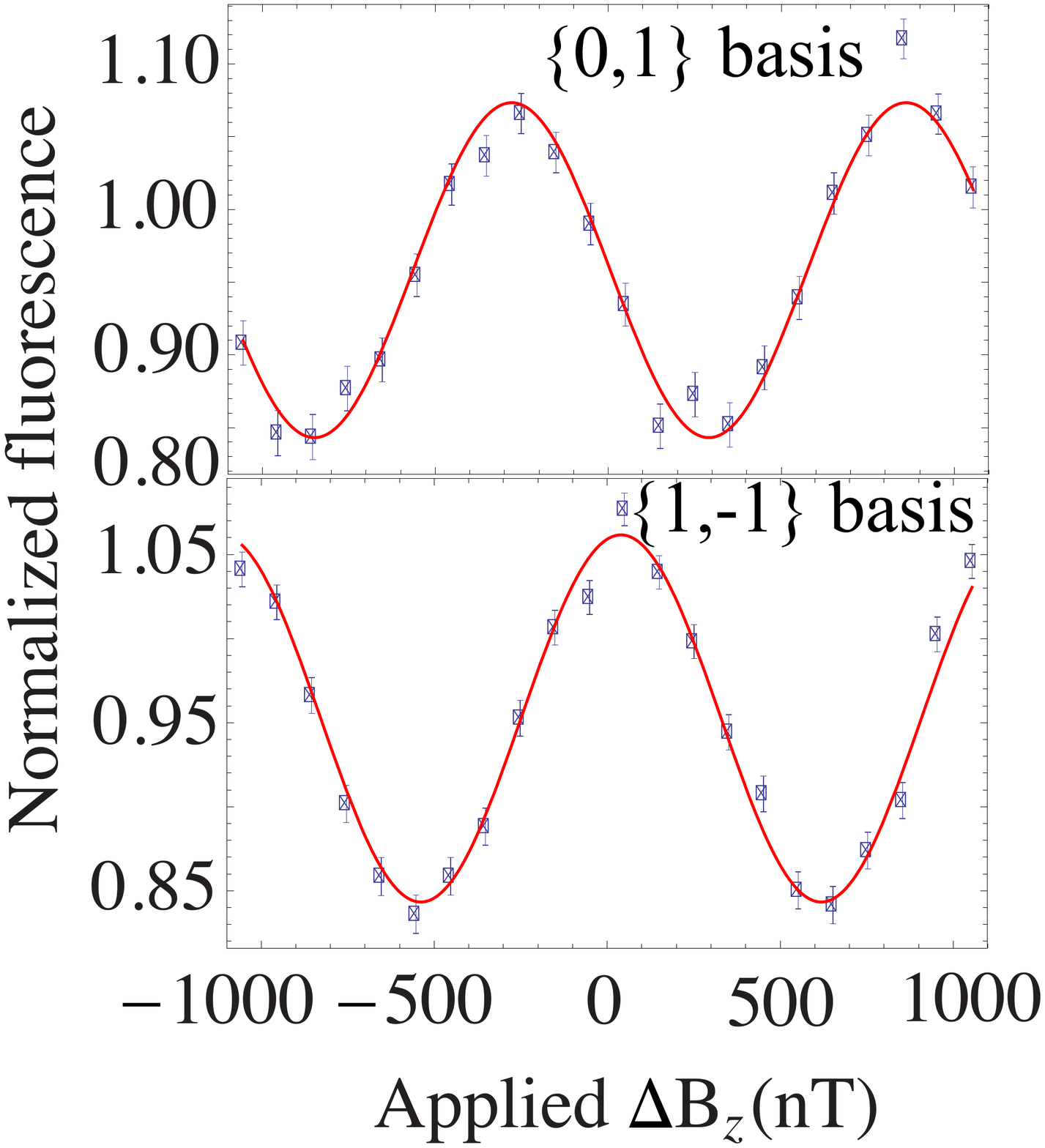,clip=1,width=0.9\linewidth,angle=0}\vspace{0cm}\hspace{0.3cm}
\caption{ Magnetometer calibration curves for the $\{0,1\}$ basis (up) and $\{1,-1\}$ basis (bottom),respectively. }
\label{calibrationcurve}
\end{figure}

Due to finite photon collection efficiency and imperfect spin-state readout, we do not actually realize the spin-projection-noise limited sensitivity given in Eq. \eqref{stvt}. Instead, the experimental sensitivity is limited by the photon shot noise. We estimate the sensitivity by keeping the free precession interval, $\tau\approx T_2^\star/2$, constant and varying the applied longitudinal magnetic field, $B_z$. The calibration of $B_z$ as a function of the current applied to our Helmholtz coil is determined by varying the current and monitoring peaks in the Ramsey spectra similar to the inset of Fig. 1(d). The resulting fluorescence signal is plotted in Fig. \ref{calibrationcurve} for both $\{1,\mbox{-}1\}$ and $\{0,1\}$ bases.

The curves are well described by the function
\be P(B_z)=\alpha \frac{T}{\tau}[1+R\sin{(2\pi\Delta m_s g_e \mu_B B_z\tau+\theta)}], \ee
where $\theta$ is a fitted phase, $2R$ is the measurement contrast, $\alpha$ is the mean number of photons collected during a single readout pulse, and $T$ is the total measurement duration. The photon shot noise for a given measurement is $\sqrt{P(B_z)}$, and therefore the photon-shot-noise-limited sensitivity is $\eta_{ph}=\frac{\sqrt{P(B_z)}}{|{dP/dB_z}|}\sqrt{T}$. The sensitivity is maximized for $2\pi\Delta m_s g_e \mu_B B_z\tau+\theta=m\pi$, where $m$ is an integer, giving
\be \eta_{ph}=\frac{1}{2\pi\Delta m_s g_e\mu_B \sqrt{\tau}}\frac{1}{R\sqrt{\alpha}}.\ee
Compared to Eq. \eqref{stvt} of the main text, we find the relation $\eta_{ph}=\eta_{min}/(R\sqrt{\alpha})$. Typical values for our magnetometer operating near the ideal case of $\tau\approx T_2^\star/2$ (Fig. \ref{calibrationcurve}) are $2R\approx0.25$ and $\alpha\approx0.01$.

\subsection{Allan deviation}
To study the Allan deviation of the measured noise, we used the following formula, \be \sigma_B(\tau)=\sqrt{0.5\overline{(B_\tau((n+1)\tau)-B_\tau(n\tau))^2}}, \ee where $\tau$ is the gate time, $\overline{\cdots}$ means average over all integer $n$, and $B_\tau(n\tau)$ is the mean value of the measured magnetic field for the gate time in the range $n\tau<t<(n+1)\tau$.

\end{document}